# Lorentz transformation, time dilation, length contraction and Doppler Effect - all at once

Bernhard Rothenstein[1] and Stefan Popescu[2]

1) Politehnica University of Timisoara, Physics Department,
Timisoara, Romania brothenstein@gmail.com
2) Siemens AG, Erlangen, Germany stefan.popescu@siemens.com

**Abstract.** *We present a simple derivation of the Lorentz transformations for the space-time coordinates of the same event. It is based on the relative character of length and time interval as measured by observes in relative motion. We begin by accepting that the relative motion modifies in some particular way the result of these measurements. However we do not postulate the character of this distortion i.e. whatever it is dilation or contraction. The formulas accounting for length contraction, time dilation and Doppler shift are a byproduct of this derivation.*

### 1. Introduction

Time and time intervals, space coordinates and distances are fundamental concepts in physics. They can be absolute or relative having the same magnitude or different magnitudes for inertial observers in relative motion. Classical physics solves the problem by stating that the time interval has an absolute character being the same for all inertial observers in relative motion. Forced by experiments, special relativity theory abandons the concept of absolute time by considering that the same time interval has different magnitudes for different inertial observers in relative motion. The length also has an absolute character in Newton's mechanics. Similarly to time, in order to account for well tested experiments, the length gets a relative character in special relativity theory.

We involve in our derivation two inertial reference frames I(XOY) and I'(X'O'Y'). Following an one space dimensions approach we consider only events that take place on the overlapped OX(O'X') axes. The corresponding axes of the two frames are parallel to each other with the OX(O'X') axes overlapped. At the origin of time (t=t'=0) the origins O and O' are located at the same point in space. Frame I'(X'O'Y') moves with constant speed V relative to I(XOY) in the positive direction of the overlapped OX(O'X') axes. At each point M(x) along the OX axis we find a clock $C(x)$. In order to be operational, all these clocks should display the same running time, a condition that can be fulfilled by performing the clock synchronization procedure proposed by Einstein. This proceeds as follows: the clock we should synchronize in I(XOY) are $C_0(0)$ located at its origin O and the clocks $C(x)$. Clock $C_0(0)$ located at the origin O is ticking, whereas clock $C(x)$ is stopped and fixed to read t=x/c which is the time required for a light signal started at t=0 from origin O to arrive at the $C(x)$ location. At the moment when clock $C_0(0)$ reads t=0 a source of light S(0) located at the origin O emits a light signal in the positive direction of the OX axis. Once arrived at clock $C(x)$ the light signal starts this clock and from now on the two clocks will display the same running time.

A similar clock synchronization procedure takes place in I'. There a clock $C'_0(0)$ located at the origin O' is ticking. A clock $C'(x')$ which should be synchronized with clock $C'_0(0)$ is stopped and fixed to read t'=x'/c, the time during which a light signal started at t'=0 from origin O' arrives at $C'(x')$ location. From now the synchronization procedure continues as in I. The concept of event is defined as a physical occurrence taking place at a given point in space M and at a give time t. This event is characterized by the space coordinate x of the point M where it takes place and by the reading t of the clock $C(x)$ located at the point M when the event takes place. The notation E(x,t) defines an event that takes place at the point M(x) when the clock $C(x)$ located at that point reads t. E'(x',t') defines an event detected from frame I' that takes place at a point M'(x') when the clock $C'(x')$ located at that point reads t'. Relativists consider that events E(x,t) and E'(x',t') represent the same event if they take place at the same point in space when the clocks $C(x)$ and $C'(x')$ located at that point read t and respectively t'.

The fundamental problem in special relativity theory is to find a relationship between the space-time coordinates of the same events. The formulas which perform this task are the Lorentz-Einstein transformations (LET) for the space-time coordinates of the same events.

Reviewing the different approaches to special relativity theory we identify two different trends. Peres[1] for example derives directly from the relativity postulates the formulas which account for the fundamental relativistic effects like time dilation, length contraction, Doppler Effect and addition law of parallel velocities. He doesn't make use at all of the Lorentz-Einstein transformation equations, considering that using them will obscure the physics behinds. Others authors first derive the LET and then use their equations as basis to derive the formulas which account for the relativistic effects mentioned above[2]. In our complementary and simpler approach we start by acknowledging the relative character of length and time interval measurements. However we do not postulate the absolute character of the distortions introduced by the relative motion of the observes (i.e. whatever it is dilation or contraction). Our derivation is reduced to the addition and comparison of lengths measured by observers of the same inertial reference frame. Levy[3] and other authors before him[4] follow a similar approach but they first derive the formulas which account for the time dilation and length contraction, using the latter to add lengths measured by observers of the same inertial reference frame. The way in which the Lorentz transformations are presented in textbooks is presented by Ziegler[5].

When speaking about time intervals, relativists make a net distinction between the proper time interval and the distorted (improper) time interval. The proper time interval between two events is measured as a difference between the readings of the same clock when the involved events take place in front of it. The improper (distorted) time interval is measured as a difference between the readings of two distant and synchronized clocks of the same inertial reference frame when the involved events take place in front of them respectively. When speaking about lengths, relativists make a net distinction between the concept of the proper length of a rod measured by an observer relative to whom it is in a state of rest and its distorted length measured by an observer relative to whom it moves with constant speed. In order to illustrate the difference between the two concepts of time consider that observers from I measure the speed of clock $C'_0(0)$. They use a meter stick at rest in I(XOY) located along the OX axis with its left end at the origin O. At t=0 the clock $C'_0(0)$ is located in front of clock $C_0(0)$ and both clocks display a zero time. After a given time of motion, clock $C'_0(0)$ arrives at the right end of the meter stick where a clock $C(L_0)$ reads t, while the moving clock reads t'. By definition the speed of the moving clock is

$$V = \frac{L_0}{t-0} = \frac{L_0}{\Delta t}. \tag{1}$$

The time interval Δt=(t-0) is measured as a difference between two readings of two different clocks (the reading t of clock $C(L_0)$ when the moving clock arrives at its location and the reading t=0 of clock $C_0(0)$ when the moving clock starts its trip from this location). Measured under such conditions Δt=t-0 represents an improper time interval. The time interval (Δt')$_0$=(t'-0) is measured as a difference between two readings of the same clock $C'_0(0)$ (the reading t' of the moving clock when it arrives at the right end of the meter stick t' and its reading t'=0 when it started its trip from the left end). Measured under such conditions (Δt')$_0$ represents a proper time interval. Relativists consider that clock $C'_0(0)$ commoving with observer $R'_0(0)$ represents his wrist watch.

Let consider that observer $R'_0(0)$ uses his wrist watch $C'_0(0)$ to measure the speed of the meter stick which moves relative to him with speed V in the negative direction of the overlapped axes. Under such conditions $R'_0$ measures a distorted length L of the moving rod. The wrist watch of the observer reads t'=0 when the left end of the rod passes in front of him and t' when the right end of the rod arrives at his location. The observer measures the proper time interval (Δt')$_0$=(t'-0). By definition the speed of the rod relative to I'(X'O'Y') is

$$V = \frac{L}{t'-0} = \frac{L}{(\Delta t')_0}. \tag{2}$$

Combining (1) and (2) results in

$$\frac{L_0}{(\Delta t)} = \frac{L}{(\Delta t')_0} \tag{3}$$

or

$$(\Delta t)L = L_0(\Delta t')_0. \tag{4}$$

Because proper physical quantities have the same magnitude for all inertial observers in relative motion, their product is a relativistic invariant, as is the product between the distorted length and the distorted time interval. This enables us to conclude that if the length contracts then the time dilates and vice versa. At this point of our derivation it is advisable to state the first postulate of SRT according to which "the laws of physics are the same in all inertial reference frames in uniform relative motion" or equivalently: "by performing experiments confined in an inertial reference frame it is impossible to detect whether the reference frame is at rest or in a state of uniform rectilinear motion". The following postulates are of use:[6]

    a. Each LT must be a single valued function of all its arguments. If the arguments are the space-time coordinates of the same events, they should interfere in the LT at the first power.

    b. Reciprocal space-time measurements of similar meter sticks and clocks at rest in two different inertial reference frames I and I' by observers at rest in I' and respectively in I yield identical results.

There are several types of transformation equations relating time intervals. We distinguish between the following possible cases:
- Relating a proper time interval in one reference frame to an improper time interval in other reference frame.
- Relating two improper time intervals measured in two inertial reference frames.
- Relating two proper time intervals measured in two inertial reference frames.

In what concerns the lengths, the transformation equation can relate a proper length measured in one of the reference frame to a distorted length measured in the other inertial reference frame or a proper length measured in one reference frame to a proper length measured in the other one.

Concentrating our analysis in the case when observers from I' measure a proper time interval $(\Delta t)_0$ and observers from I measure an distorted time interval $\Delta t$, in accordance with (a) these results should be related by

$$\Delta t = a_t (\Delta t)_0 \quad (5)$$

where $a_t$ represents a transformation factor which depends on the relative velocity V, but not on $(\Delta t)_0$. In accordance with (b), if observers from I measure a proper time interval in I and observers from I' measure an improper time interval then the two time intervals should be related by the same factor $a_t$. In the case when observers from I' measure a proper length $L_0$ and observers from I measure a distorted length L the two lengths are related by

$$L = a_L L_0 \quad (6)$$

where the transformation factor $a_L$ depends on the relative speed, but not on $L_0$. Combining (5) and (6) results in

$$a_L a_t = 1. \quad (7)$$

An important consequence of the fact that the clocks are synchronized à la Einstein in I and in I' is that the space-time coordinates of an event generated by the synchronizing light signal are related by

$$(x-0) = c(t-0) \quad (8)$$

and by

$$x' - 0 = c(t' - 0). \quad (9)$$

**2. Lorentz-Einstein transformations for the space-time coordinates of the same event, length contraction, time dilation, Doppler Effect and addition law of parallel speeds**

Consider the relative position of the reference frames I and I' as shown in Figure 1 as detected from I when the clocks of this frame read t.

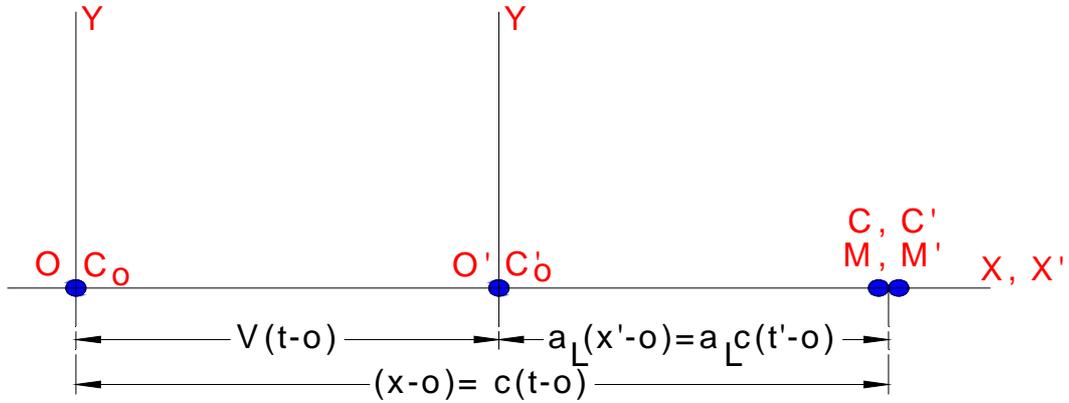

*Figure 1. The relative position of the reference frames I and I' as detected from I when the clocks of that frame read t. Only lengths measured by observers from I are marked.*

As a result of the relative character of length we can mark on it only distances measured by observers from I. The distance between the origins O and O' is V(t-0). A point M(x) is located at a proper distance from O equal to Δx=x-0=c(t-0). The point M'(x') located at the same point in space as point M(x) is located at a proper distance Δx'=x'-0=c(t'-0) from O'. Measured by observers from I this becomes the distorted length $a_L$(x'-0)=c(t'-0). At the points M(x) and M'(x') located at the same point in space we find the clocks $C$(x) of I and $C'$(x') of I' reading t and respectively t'. Adding only lengths measured from frame I as depicted by Figure 1 we have:

$$x - 0 = V(t - 0) + a_L(x' - 0) \qquad (10)$$

or

$$a_L c(t' - 0) = ct(1 - \frac{V}{c}). \qquad (11)$$

Figure 2 illustrates the relative position of the reference frames I and I' as detected from I' when the clocks of that frame read t'.

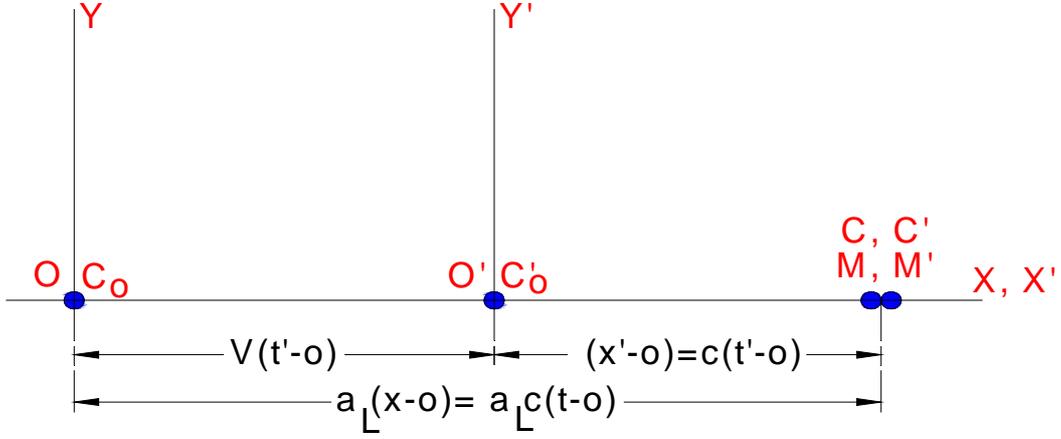

*Figure 2. The relative position of the reference frames I and I' as detected from I' when the clocks of that frame read t'. Only lengths measured by observers from I' are marked.*

The distance between the origins O and O' is Vt' and the point M'(x') is located at a proper distance (x'-0)=c(t'-0) from O'. The proper distance (x-0)=c(t-0) measured in I becomes the distorted length $a_L(x-0)=a_L c(t-0)$. Adding only lengths measured in I' yields

$$a_L(x-0) = V(t'-0) + (x'-0) \qquad (12)$$

or

$$a_L c(t-0) = ct'(1+\frac{V}{c}). \qquad (13)$$

Multiplying (11) and (13) side by side we obtain

$$a_L = \sqrt{1-\frac{V^2}{c^2}} \qquad (14)$$

with (10) leading to

$$x'-0 = \frac{(x-0)-V(t-0)}{\sqrt{1-\frac{V^2}{c^2}}} \qquad (15)$$

whereas (12) leads to

$$x = \frac{V(t'-0)+(x'-0)}{\sqrt{1-\frac{V^2}{c^2}}}. \qquad (16)$$

Equations (15) and (16) can be presented also as

$$x' = \frac{x - Vt}{\sqrt{1 - \frac{V^2}{c^2}}} \tag{17}$$

and as

$$x = \frac{x' + Vt'}{\sqrt{1 - \frac{V^2}{c^2}}}. \tag{18}$$

Solving (17) and (18) for t and t' we obtain

$$t' = \frac{t - \frac{V}{c^2} x'}{\sqrt{1 - \frac{V^2}{c^2}}} \tag{19}$$

and

$$t = \frac{t' + \frac{V}{c^2} x'}{\sqrt{1 - \frac{V^2}{c^2}}}. \tag{20}$$

Equations (17), (18), (19) and (20) represent the Lorentz-Einstein transformations.

In order to be able to properly use the transformation equations derived above, it is important to have a correct representation of the physical quantities they relate. Consider that at the points M(x) and M'(x') located at the same point in space as detected from I(XOY) and respectively I'(X'O'Y') we find the clocks *C*(x) and *C'*(x') reading t and respectively t'. The event E(x,t) is associated with the fact that the clock C(x) reading t is located at the point M(x). The event E'(x',t') is associated with the fact that the clock *C'*(x') reading t' is located at the point M'(x'). By definition the two events represent the same event with the transformation equations derived above relating their space-time coordinates. The distances (x-0) and (x'-0) represent proper lengths, whereas the time intervals (t-0) and (t'-0) represent improper time intervals.

The derivations performed above have as a byproduct the formulas that account for the time dilation and the length contraction effects. If we consider that in one of the involved reference frames, say I', observers measure the proper length $L_0$ while observers from I relative to whom it moves measure its distorted length related in accordance with (6) as

$$L = \sqrt{1 - \frac{V^2}{c^2}} L_0. \tag{21}$$

Because $L < L_0$ relativists say that a length contraction effect takes place. Consider that observers from say I' measure a proper time interval $\Delta t$ while observers from I measure an improper time interval $(\Delta t)_0$. In accordance with (5) and (7) the two time intervals are related by

$$\Delta t = \frac{(\Delta t)_0}{\sqrt{1 - \frac{V^2}{c^2}}}. \tag{22}$$

The problem is now to find an equation that relates two proper time intervals measured in I and respectively in I'. Consider that an observer $R_0(0)$ of I located at O and an observer $R'_0(0)$ of I' located at O' are both equipped with light sources provided with shutters. Observer $R_0$ opens the shutter for a proper time interval $\Delta t=(t-0)$ in order to generate a "light rod" of length $\Delta x=c(t-0)$. Opening the shutter of his light source for a proper time interval $(\Delta t')_0 = c(t'-0)$, observer $R'_0(0)$ generates a "light rod" of length $(\Delta x')=c(t'-0)=(\Delta t')_0$. Adding only lengths measured by observers from I we obtain

$$c(\Delta t)_0 = a_L c(\Delta t')_0 (1+\frac{V}{c}) \tag{23}$$

with the result that the proper time intervals $(\Delta t)_0$ and $(\Delta t')_0$ are related by

$$(\Delta t)_0 = (\Delta t')_0 \sqrt{\frac{1+\frac{V}{c}}{1-\frac{V}{c}}}. \tag{24}$$

Adding only lengths measured in I' we obtain

$$a_L c(\Delta t')_0 = c(\Delta t) - Vt \tag{25}$$

or

$$(\Delta t')_0 = \sqrt{\frac{1-\frac{V}{c}}{1+\frac{V}{c}}} (\Delta t)_0. \tag{26}$$

The proper time intervals involved in the transformation equations derived above could have the following physical meanings:

– $(\Delta t)_0$ and $(\Delta t')_0$ represent the proper time intervals for which the observers $R_0$ and $R'_0(0)$ should maintain the shutters open in order to generate the events E(x=ct, t=x/c) in I and E'(x'=ct, t'=x'/c) in I'. (Radar detection).

– $(\Delta t)_0$ represents the proper period at which the stationary observer $R_0(0)$ emits successive light signals in the positive direction of the OX axis while $(\Delta t')_0$ represents the proper period at which the receding observer $R'_0(0)$ receives them. In this case equations (25) and (26) could be converted into emission and reception frequencies $\nu_e = \frac{1}{(\Delta t)_0}$ and $\nu_r = \frac{1}{(\Delta t')_0}$ accounting for the Doppler Effect in the optical domain with stationary source of light and receding observer.

**Conclusions**

The Lorentz transformations for the space-time coordinates presented above present the advantage that it has as a byproduct the formulas which account for the basic relativistic effects, used by many Authors as a starting point in their approach to the same problem. During the derivations we explain the meaning of the physical quantities involved in the transformation process making them transparent.

**References**

[1] Asher Peres, "Relativistic telemetry", Am.J.Phys, **55**, 516-519 (1987)

[2] Robert Resnick, *Introduction to Special Relativity,* (John Wiley and Sons. Inc. New York, Sydney, 1968) pp.56-62

[3] J.-M. Levy, "A simple derivation of the Lorentz transformation and of the accompanying velocity and acceleration changes", Am.J.Phys. **75,** 615-618 (2007)

[4] A.V. Astahov, *Mechanics. Kinetic Theory of Matter* (Glav.Red. Fiz-Mat Lit. Moscow 1977) pp. 49-53 (In Russian)

[5] Alfred Ziegler, "The role of the two postulates of special relativity", arXiv:0708.0988v17, Aug 2007

[6] J.H. Field, "A new kinematical derivation of the Lorentz transformation and the particle description of light", Helv.Phys.Acta, **70,** 542-564 (1997)